\documentclass[letterpaper,twocolumn,aps]{revtex4}

\usepackage{graphicx}

\begin{document}

\title{Fermi surface evolution and Luttinger theorem in Na$_x$CoO$_2$: a systematic photoemission study}

\author{H.-B. Yang,$^{1}$ Z.-H. Pan,$^{1}$ A.K.P. Sekharan,$^{1}$ T. Sato,$^{2}$ S. Souma,$^{2}$ T. Takahashi,$^{2}$ 
R. Jin,$^{3}$ B.C. Sales,$^{3}$ D. Mandrus,$^{3}$  A.V. Fedorov,$^{4}$ Z. Wang,$^{1}$ and H. Ding$^{1}$}

\affiliation{
(1) Department of Physics, Boston College, Chestnut Hill, MA 02467 \\
(2) Department of Physics, Tohoku University, 980-8578 Sendai, Japan \\
(3) Condensed Matter Science Division, Oak Ridge National Laboratory, Oak Ridge, TN 37831\\
(4) Advanced Light Source, Lawrence Berkeley National Laboratory, Berkeley, CA 94720
}
\begin{abstract}
\noindent 
We report a systematic angle-resolved photoemission study on Na$_x$CoO$_2$ for a wide range of Na concentrations 
($0.3 \leq x \leq 0.72$). In all the metallic samples at different $x$, we observed (i) only a single hole-like Fermi 
surface centered around $\Gamma$ and (ii) its area changes with $x$ according to the Luttinger 
theorem. We also observed a surface state that exhibits a larger Fermi surface area. 
The $e_g^\prime$ band and the associated small Fermi surface pockets near the $K$ points predicted by band calculations 
are found to ``sink'' below the Fermi energy in a manner almost independent of the doping and temperature.
\noindent 

\end{abstract}
\maketitle
The surprising discovery of superconductivity on Na$_x$CoO$_{2} \cdot y$H$_2$O \cite{Discovery}
raises many interesting questions on the nature of pairing and
its connection to the high-$T_c$ superconductivity. 
The phase diagram of the cobaltate Na$_x$CoO$_{2} \cdot y$H$_2$O, with varying electron doping $x$ and water intercalation $y$ 
over a wide range, is very rich and complicated. In addition to superconductivity, it exhibits charge order, magnetic 
order including a metamagnetic transition, and other structural transitions \cite{Foo,Luo,Sales,Huang}. 
The physics of these phases and the transitions among them is of importance by itself, and offers an excellent platform for 
studying correlated triangular lattice fermion systems. 

Understanding the evolution of the low-energy electronic structure
such as the quasiparticle (QP) dispersion and the Fermi surface (FS) topology is a necessary step toward
understanding the diverse physical properties and the nature of the pairing interaction in this class
of transition metal oxide. It is therefore important to gain the precise knowledge of the FS structure and 
the low-energy excitations in the unhydrated Na$_x$CoO$_{2}$ which have been the focus in many recent theoretical 
and experimental efforts. The cobaltate is a multi-orbital system where the Co$^{4+}$ is in the 3d$^5$ 
configuration, occupying the lower $t_{2g}$ band complex similar to the ruthenate Sr$_2$RuO$_4$.
First principle band calculations have predicted that Na$_x$CoO$_{2}$ has 
a large FS associated with the $a_{1g}$ band centered around the $\Gamma$ point and 
six small FS pockets of mostly $e_g^\prime$ character near the $K$ points for a wide range
of $x$ \cite{Singh_FS,KWLee}. Based on this band structure, it has been proposed that the large density of states 
contribution from the six FS pockets and the nesting condition among them enhance the spin fluctuations and
lead to superconducting pairing of the QPs on these FSs \cite{Johannes}.
However, angle-resolved photoelectron spectroscopy (ARPES) measurements on the cobaltate with high Na concentrations 
($x \sim$ 0.6 - 0.7) revealed only the large FS \cite{Hasan,HBYang}. It was also noticed that the enclosed FS area, i.e.
the density of holes, may not satisfy the Luttinger theorem which is a fundamental QP counting rule 
in interacting electron systems. It is therefore desirable to study the evolution of the QP
band dispersion and the FS, especially the fate of the FS pockets as a function of the Na concentration $x$.
This is the focus of this work. We choose a set of metallic Na$_x$CoO$_2$ with $0.3 \leq x \leq 0.72$.
The insulating phase at $x=0.5$ due to Na dopant order and the magnetic phase at $x \geq 0.75$ are outside the scope of 
this study. 

High quality single crystal Na$_x$CoO$_2$ samples were prepared by the flux method and
subsequent sodium deintercalation. 
ARPES experiments were performed at the Synchrotron Radiation Center, WI, and the 
Advanced Light Source, CA. High-resolution undulator beamlines and Scienta analyzers with a capability of multi-angle 
detection have been used. Most spectra were measured using 100 eV photons which we found to suit well for the FS
mapping of Co 3$d$ orbitals. However, some spectral features are better revealed at lower photon energies which will be 
discussed below. The energy resolution is $\sim$ 10 - 40 meV, and the momentum resolution $\sim0.02$ \AA$^{-1}$. 
Samples are cleaved and measured \emph{in situ} in a vacuum better than $8\times10^{-11}$ $Torr$ at low temperatures 
(20 - 40 K) on a flat (001) surface.  

\begin{figure}[{here}]
\includegraphics[ width = 8cm]{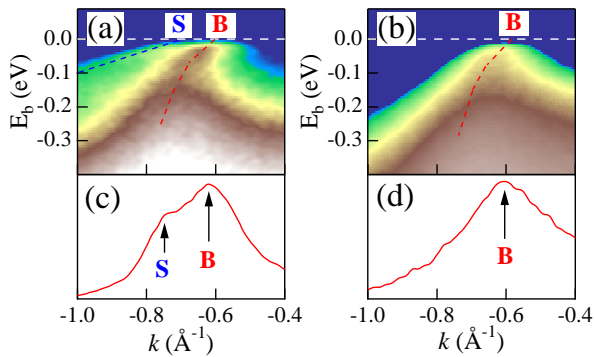}
\vspace{-10pt}
\caption{ARPES $E$-$k$ intensity plots at $T$ = 40 K on (a) a fresh surface of Na$_{0.72}$CoO$_2$ and (b) an aged surface 
of the same sample after a thermal cycle to 200 K. Red dashed lines are guides for the bulk band (marked as ``B''), and the 
blue dashed line in (a) traces the surface state (marked as ``S''). The lower panel shows the momentum distribution 
curves at $E_F$ obtained on (c) the fresh surface and (d) the aged surface where the surface state vanishes. 
}
\label{VB}
\vspace{-17pt}
\end{figure}

Due to the surface-sensitive nature of ARPES, extra care needs to be taken in assessing how much ARPES data 
represent the bulk electronic structure. In many cases, surface states arise from surface reconstruction. One useful
way to distinguish the surface state from the bulk state is to utilize the property that the surface state 
is more sensitive to surface disorder. 
The technique used in the ARPES community is to thermally cycle a sample and thus introduce disorder to the surface. 
This method is often found to be very effective in removing the surface state in Sr$_2$RuO$_4$ 
\cite{Damascelli,SCWang}. On a freshly cleaved Na$_x$CoO$_2$ surface, we often observed an unusual double-band 
dispersion \cite{HBYang}, as shown in Fig.~1(a). However, after the thermal cycle (40 K - 200 K - 40 K), one of the two 
bands (labeled as ``S'' in Fig.~1) disappears, which is likely a surface state. We note that this surface state has a larger 
Fermi vector ($k_F$) and a smaller Fermi velocity ($v_F$). The ``surviving'' band (labeled as ``B'') is likely bulk-representative. 
Throughout the paper, we will only discuss the bulk-like band structure. The bulk nature of the ARPES data reported 
here is further verified by our x-ray ARPES which has much longer escape length for photoelectrons. 
The reason we use lower photon energy here is that it has much better 
resolutions in both energy and momentum, which enable us to study the FS and low-energy excitations 
more precisely. 

We start with a survey of the valence bands and its doping dependence, as shown in Fig.~2. Panels (a) and (b) 
show the second derivative intensity (SDI) plots (a common practice to display dispersion for broad peaks) along 
$\Gamma$-$M$ for $x$ = 0.48 and 0.72, respectively. There are four major band branches, with the one at 
$\sim$ 0.5 eV being assigned to the Co 3$d$ $t_{2g}$ orbitals, and the other three at $\sim$ 3, 4.5, 6 eV to the 
O 2$p$ orbitals, according to local density approximation (LDA) calculation \cite{Singh_FS}. The comparison between 
the two doping levels, plotted in Fig.~2(c), shows a clear and non-trivial doping dependence on the energy level of 
these bands. For the Co 3$d$ orbitals, the average energy difference between the two doping levels is $\sim$ 0.1 eV, 
which is likely due to a shift of the chemical potential caused by the change of carrier numbers. However, for the O 2$p$ 
orbitals, the energy shifts ($\sim$ 0.5 eV) between the two doping levels are larger. 
We note that this doping dependence of the O-2$p$ orbitals can explain well the doping-dependent shift of the 
$\alpha$ peak in optical conductivity measurement which is associated with the transition from the fully 
occupied O-$p$ states to the unoccupied Co $e_g$ states \cite{NLWang2}. 

\begin{figure}[{here}]

\includegraphics[ width = 8cm]{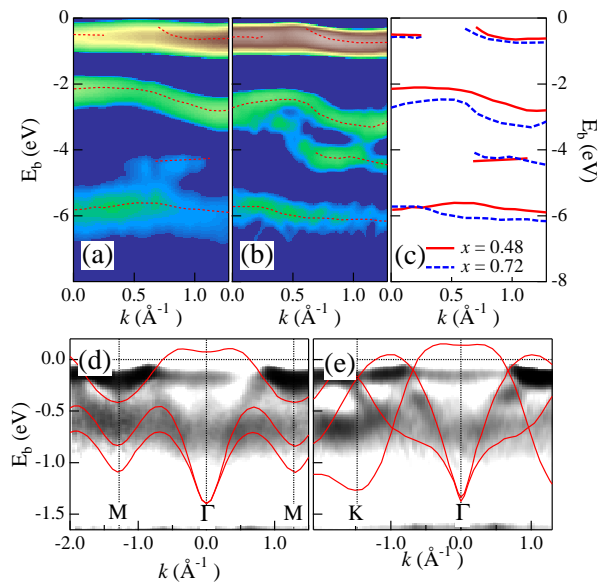}
\vspace{-10pt}
\caption{SDI plots of the valence bands along $\Gamma$-$M$ for (a) $x$ = 0.48 and (b)
$x$ = 0.72. Red dashed lines are the guides for band dispersion. (c) Comparison of Co-3$d$ and O-2$p$ orbitals 
between $x=0.48$ and $x=0.72$. The lower panel shows the SDI plots for the Co-3$d$ orbitals 
in Na$_{0.35}$CoO$_2$ along (d) $\Gamma$-$M$ and (e) $\Gamma$-$K$. 
Red lines are band dispersions from the LDA calculation \cite{KWLee}.
}
\label{VB}
\vspace{-17pt}
\end{figure}

Zooming in on the Co-3$d$ orbitals, one can identify more dispersive features. Figs.~2 (d) and (e) display the 
SDI plots of Na$_{0.35}$CoO$_2$ on a smaller energy scale ($\sim$ 1.5 eV) along $\Gamma$-$M$ 
and $\Gamma$-$K$ directions, respectively. By comparing to the LDA band structure for the same Na content \cite{KWLee}, 
one can assign orbital characters to the dispersive features. However, the measured occupied 
bandwidths ($\sim$ 0.7 - 0.8 eV) are, remarkably, narrower than the calculated ones (1.3 - 1.4 eV), 
resulting in approximately a factor of two in the overall bandwidth reduction, 
indicating the importance of strong electron correlations in this material. 
Note that the flat features at the unoccupied wavevectors ($k$) visible in Figs.~2(d) and (e) near the Fermi energy ($E_F$)
may be due to the incoherent background often observed in correlated materials.

To study the low-energy excitations and the FS more closely, we further zoom in near $E_F$ 
(within 0.2 eV), as shown in Fig.~3 where a sample of Na$_{0.48}$CoO$_2$ is measured. This time we directly display 
the intensity plots since the near-$E_F$ features are much sharper (see the energy distribution curves (EDCs) in Fig.~3(a)), 
and the SDI plot cannot 
reveal the near-$E_F$ band well due to the interference of the Fermi-Dirac function. During the ARPES measurement, we 
took many parallel cuts (five of them are shown in Figs.~3 (b) - (f)) in $k$ space. It is important to 
have such long cuts that cover several Brillouin zones (BZs) in order to accurately determine the FS, 
since they can help eliminate many potential problems such as sample misalignment and matrix element effects. 
Indeed, we have a small misalignment of about $3^\circ$ in this sample as can be seen in Fig.~3(g). Note that the intensity 
is much weaker in the 2nd BZ due to the photoemission matrix element effect. The FS is determined by plotting 
the ARPES intensity within a narrow energy window ($\pm$2 meV) at $E_F$ in the two-dimensional (2D) $k$ space
\begin{figure}[{here}]
\includegraphics[width = 8cm]{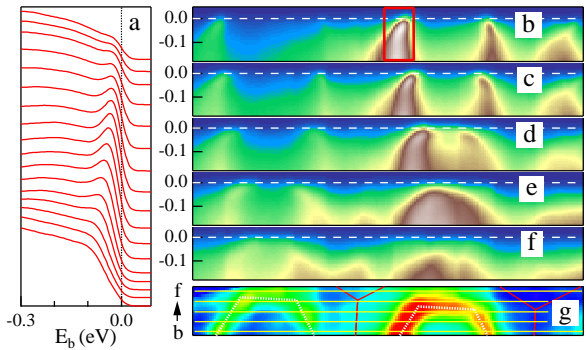}
\vspace{-10pt}
\caption{
An example of the FS mapping on Na$_{0.48}$CoO$_2$. (a) Representative EDCs within the red box in panel (b). (b) - (f) $E$-$k$ intensity plots along the long cuts 
indicated by the yellow solid lines in panel (g) where FS contours obtained from the intensity at $E_F$, and BZ boundaries 
(thin red hexagons), are plotted. White dashed lines are the guides for the FSs.
}
\label{Crossing}
\vspace{-17pt}
\end{figure}
as shown in Fig.~3(g) where partial FS contours over two BZs are extracted from the five long cuts shown above.
We also note that the low-energy band is further renormalized due to a strong ``kink'' in the dispersion 
observed at the energy scale 
of 70 - 100 meV (see Figs.~3(b) - (d)). The behavior of this ``kink'' and the relationship to the well-known ``kink'' 
in the dispersion observed in high-$T_c$ cuprates \cite{Valla,Bogdanov,Kaminski,Lanzara}
will be discussed in a separate publication.

The FS we observed, as shown in Fig.~3(g), corresponds to the large $a_{1g}$ FS centered at the $\Gamma$ point
predicted by the LDA calculation \cite{Singh_FS}. However, the six small FS pockets associated with the
$e^\prime_{g}$ band predicted by the LDA are not present in our measurements for the entire range of $0.3\le x\le 0.72$. 
Instead, we observed a broad peak that approaches but never reaches $E_F$ near the
$K$ points, as indicated by the black triangles in Fig.~4(a). In addition, we observed two more dispersive features. 
The one indicated by the green triangles belongs to the $a_{1g}$ band, and the other indicated by the blue triangles is 
assigned to the other $e^\prime_{g}$ band. The reason that the intensity of the 
$a_{1g}$ band is much weaker in Fig.~4 than in Fig.~3 is that different photon energies were used in the two 
measurements. While we used 100-eV photons in Fig.~3 to enhance the Co-3$d$ character, we found that the 
$e^\prime_{g}$ band near the $K$ points is greatly enhanced at lower photon energies (30 eV in Fig.~4). This may 
suggest a stronger mixing of O-2$p$ character in the $e^\prime_{g}$ bands since the O $p$ orbitals are more sensitive 
to the low-energy photons, a property often observed in the ARPES studies of cuprates. 

\begin{figure}[{top}]
\includegraphics[width = 8cm]{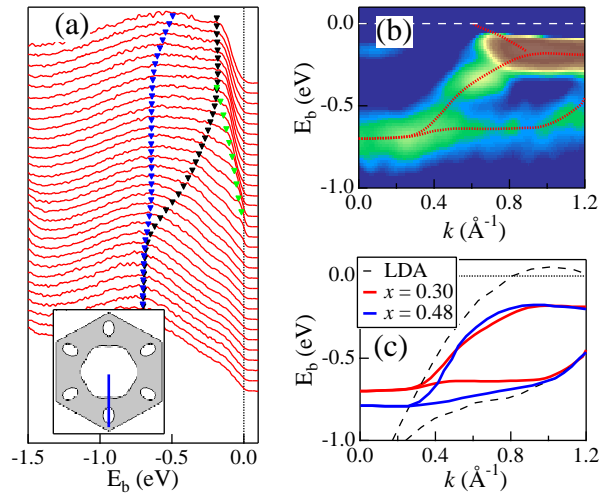}
\vspace{-10pt}
\caption{
 ``Sinking'' pockets near the $K$ points. (a) EDCs of Na$_{0.3}$CoO$_2$ along $\Gamma$-$K$ 
 measured using 30 eV photons. Triangular markers are the guides for the band dispersion. The insert shows the location of 
 the cut in the BZ. (b) SDI plot for the same measurement. Red dashed lines, representing the 
 triangular markers in panel (a), are the guides for the dispersion. (c) A comparison of the measured $e_{g}'$ bands at 
 $x$ = 0.3 (red solid lines) and 0.48 (blue solid lines) to the LDA calculation \cite{KWLee} 
(black dashed lines).
}
\label{FS}
\vspace{-17pt}
\end{figure}

The broad $e^\prime_{g}$ bands can be observed more clearly in the SDI plot, as shown in 
Fig.~4(b). In comparison, the measured dispersions are reminiscent of the $e^\prime_{g}$ bands 
calculated by the LDA \cite{Singh_FS,KWLee}, as shown in Fig.~4(c). However, important differences exist: 1) the measured 
bandwidth is narrower than the calculated one; 2) the upper branch of $e^\prime_{g}$ bands does not cross the 
Fermi level - the predicted FS pockets ``sink'' below $E_F$. More significantly, this ``sinking'' pocket does \emph{not} 
have doping dependence for its band top, as shown in Fig.~4(c), while the band 
bottom shifts with doping ($\sim$ 100 meV shift between $x$ = 0.3 and $x$ = 0.48). 
This behavior reminds us of the leading edge property associated with the opening of an energy gap at the FS.
However, we did not observe clear leading-edge shift for temperatures up to 250 K. 

We summarize our ARPES results on the FS evolution in Na$_x$CoO$_2$ in Fig.~5. We have measured many 
samples and obtained well-reproducible results over a wide range of Na concentrations. Figs.~5(a) - (c) show three 
examples of the measured FS for $x$ = 0.3, 0.48, and 0.72. Clearly, a single hexagonal hole-like 
FS, centered at the $\Gamma$ point, shrinks its size as $x$ increases. A direct comparison of the FS contours 
at the four doping levels, shown in Fig.~5(d), provides more quantitative information on the FS evolution. All Fermi 
surfaces have a good hexagonal shape with parallel FS edges that can be connected by a nesting 
vector ($\vec{Q}_n$). The magnitude of $\vec{Q}_n$ is estimated to be $\sim$ 1.41, 1.40, 1.20, 1.18 \AA$^{-1}$ 
($\pm$ 0.1 \AA$^{-1}$) for $x$ = 0.3, 0.48, 0.6, 0.72. 
Interestingly, these values are close to the reciprocal lattice vectors
$\Gamma K$ (1.47 \AA$^{-1}$) and $\Gamma M$ (1.28 \AA$^{-1}$).  
From Fig.~5(d), we derive the carrier 
density from the FS area which we call the ``effective Na concentration'' $x' = 1-{2A_{FS}}/{A_{BZ}}$,
where $A_{FS}$ is the area enclosed by the 2D FS, and $A_{BZ}$ is the area of the BZ. In Fig.~5(e), 
$x'$ is plotted vs the nominal Na concentration $x$. If the Luttinger theorem is satisfied, one has
$x^\prime=x$, which is indicated by the solid line in Fig.~5(e).
Within the experimental uncertainties, Fig.~5(e) shows that $x^\prime$ tracks well the $x^\prime=x$ 
line and the Luttinger theorem is thus satisfied.
The conservation of the doped electron density is quite remarkable and is consistent with the non-existence 
of the small FS pockets. In contrast, the FS pockets near the $K$ points predicted by LDA calculations
contribute to a significant portion of the total FS area at low doping.
This helps clarify the puzzle of the FS being too large in previous ARPES results 
\cite{Hasan,HBYang}. A part of the reason for this discrepancy, besides the issue of experimental accuracy, 
is that we eliminated the surface state in this study, which is tied to a larger FS as discussed above. 
We note that there may exist a small jump in the FS area across $x$ = 0.5, as shown in Figs.~5(d) and (e). 
It is known that the metallic states of Na$_x$CoO$_2$ are different across $x$ = 0.5, with Pauli-like susceptibility 
for $x <$ 0.5 and Curie-Weiss like for $x >$ 0.5. Whether the jump is real or how it relates to the magnetic 
``transition'' is still an open question that needs further investigation. 

\begin{figure}[{top}]
\includegraphics[width = 8cm]{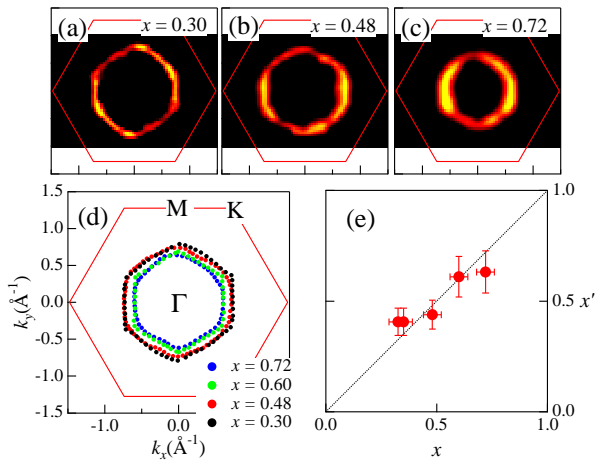}
\vspace{-10pt}
\caption{
FS evolution in Na$_{x}$CoO$_2$. (a) - (c) FSs (the intensity contours at $E_F$) for $x$ = 0.3, 
0.48, and 0.72 in the first BZ (red solid lines). (d)Overlap of the FS locations at four doping levels in the BZ. 
(e) Effective Na concentration $x'$ derived from FS area {\it v.s.} Na concentration $x$. 
The diagonal line is from the Luttinger theorem.
}
\label{FS}
\vspace{-17pt}
\end{figure}

The observation of the ``sinking'' pockets near the $K$ points for such a wide range of Na concentrations
is very intriguing. The qualitative discrepancy with the LDA band structure points to the importance
of electronic correlation effects and is a basic unresolved issue in understanding the physical properties of
the cobaltates. A recent theoretical work based on the local spin density approximation (LSDA)
takes into account the local Coulomb repulsion (U) using the LSDA+U approach and finds the absence of the
small FS pockets \cite{PHZhang}. However, the
disappearance of the small FS pockets in the LSDA calculation is due to the formation of a half-metal with spin-split
bands and spin polarized FSs, resulting in a FS area 
twice as large which is inconsistent with our observations.
The observed behavior that the top of the ``sinking'' 
pocket stays at the same energy while sizable energy shifts appear at the band bottom for different doping levels 
is unusual and seems to suggest either the opening of a gap or more intricate many-body effects. 
So far, the lack of the temperature dependence of the leading edge does not appear to support the gap-opening 
scenario. A better understanding of the correlation effects as well as those associated with the crystal field 
splitting and the hybridization with the O-$2p$ orbitals is clearly needed in order to understand these
unconventional electronic properties.

In conclusion, our ARPES results on Na$_x$CoO$_2$ over a wide range of Na concentrations ($0.3 \leq x \leq 0.72$) 
clearly show that there is only a single hexagonal FS centered around the $\Gamma$ point, with parallel 
edges being possibly nested. We find that the evolution of this hole-like FS obeys the Luttinger theorem. 
The small FS pockets near the $K$ points predicted by the LDA calculations are found to ``sink'' 
below $E_F$ with a distance to the Fermi sea almost independent of doping and temperature. 
These findings provide clear and detailed knowledge of the evolution of the electronic structure in Na$_x$CoO$_2$
and put constraints on the theoretical description of the superconductivity in the hydrated cobaltates.

We thank P.A. Lee, D.J. Singh for valuable 
discussions and suggestions, and S. Gorovikov, H. H\"{o}chst for technical support in synchrotron experiments,
This work is supported by NSF DMR-0353108, DOE DE-FG02-99ER45747, Petroleum 
Research Fund, and the MEXT of Japan. This work is based upon research conducted at the Synchrotron Radiation 
Center supported by NSF DMR-0084402, and at the Advanced Light Source supported by DOE DE-AC03-76SF00098. 
Oak Ridge National laboratory is managed by UT-Battelle, LLC, for DOE under contract DE-AC05-00OR22725.


\vspace{-10pt}
\begin{thebibliography}{10}
\bibitem{Discovery} K. Takada \textit{et al}., Nature \textbf{422}, 53 (2003).
\bibitem{Foo} M.L. Foo \textit{et al}., Phys. Rev. Lett. \textbf{92}, 247001 (2004).
\bibitem{Luo} J.L. Luo \textit{et al}., Phys. Rev. Lett. \textbf{93}, 187203 (2004).
\bibitem{Sales} B.C. Sales \textit{et al}., Phys. Rev. B \textbf{70},174419 (2004).
\bibitem{Huang} Q. Huang \textit{et al}., Phys. Rev. B \textbf{70}, 134115 (2004).
\bibitem{Singh_FS} D.J. Singh,  Phys. Rev. B \textbf{61}, 13397 (2000).
\bibitem{KWLee} K.-W. Lee \textit{et al}., Phys. Rev. B \textbf{70}, 045104 (2004).
\bibitem{Johannes} M.D. Johannes \textit{et al}., Phys. Rev. Lett. \textbf{93}, 097005 (2004).
\bibitem{Hasan} M.Z. Hasan \textit{et al}., Phys. Rev. Lett. \textbf{92}, 246402 (2004).
\bibitem{HBYang} H.-B. Yang \textit{et al}., Phys. Rev. Lett. \textbf{92}, 246403 (2004).
\bibitem{Damascelli} A. Damascelli \textit{et al}., Phys. Rev. Lett. \textbf{85}, 5194 (2000).
\bibitem{SCWang} S.-C. Wang \textit{et al}., Phys. Rev. Lett. \textbf{92}, 137002 (2003).
\bibitem{NLWang2} N.L. Wang \textit{et al}., Phys. Rev. Lett. \textbf{93}, 237007 (2004).
\bibitem{Valla} T. Valla \textit{et al}., Science \textbf{285}, 2110 (1999).
\bibitem{Bogdanov} P.V. Bogdanov \textit{et al}., Phys. Rev. Lett. \textbf{85}, 2581 (2000).
\bibitem{Kaminski} A. Kaminski \textit{et al}., Phys. Rev. Lett. \textbf{86}, 1070 (2001).
\bibitem{Lanzara} A. Lanzara \textit{et al}., Nature \textbf{412}, 510 (2001). 
\bibitem{PHZhang} Peihong Zhang \textit{et al}., Phys. Rev. Lett. \textbf{93}, 236402 (2004).

\end{thebibliography}
\end{document}